# Comparative Multi Fractal De-trended Fluctuation Analysis of heavy ion interactions at a few GeV to a few hundred GeV

Gopa Bhoumik, Argha Deb, Swarnapratim Bhattacharyya[*] and Dipak Ghosh

**Nuclear and Particle Physics Research Centre**
**Department of Physics**
**Jadavpur University**
**Kolkata - 700 032**
**India**

* Department of Physics, New Alipore College,
L Block, New Alipore, Kolkata 700053, India

Email: ppgopa@gmail.com

## Abstract

We have studied the multifractality of pion emission process in $^{16}$O-AgBr interactions at 2.1AGeV & 60AGeV, $^{12}$C-AgBr & $^{24}$Mg-AgBr interactions at 4.5AGeV and $^{32}$S-AgBr interactions at 200AGeV using Multifractal Detrended Fluctuation Analysis (MFDFA) method which is capable of extracting the actual multifractal property filtering out the average trend of fluctuation. The analysis revels that the pseudo rapidity distribution of the shower particles is multifractal in nature for all the interactions i.e. pion production mechanism has in built multi-scale self-similarity property. We have employed MFDFA method for randomly generated events for $^{32}$S-AgBr interactions at 200 AGeV. Comparison of expt. results with those obtained from randomly generated data set reveals that the source of multifractality in our data is the presence of long range correlation. Comparing the results obtained from different interactions, it may be concluded that strength of multifractality decreases with projectile mass for same projectile energy and for a particular projectile it increases with energy. The values of ordinary Hurst exponent suggest that there is long range correlation present in our data for all the interactions.

## Introduction

The study of correlation and fractality is an active area of research in many fields including heavy ion collisions [1-5]. Natural systems, which have irregular pattern at different scales, exhibit fractal nature. Fractals are generally classified into two categories i) Monofractal and ii) Multifractals. For monofractals scaling properties of the system is identical throughout the system on the other hand "Multifractals" are more complicated self-similar structure that consist of a number of weighted fractals with different non-integer dimensions. As the scaling properties are dissimilar in different parts of the system, multifractal systems require at least more than one scaling exponent to describe the scaling behavior of the system **[6].** Investigation of multifractality is of great importance as its origin may be associated with the presence of long range correlation in the system. Correlation study has the potential to provide information about the characteristics of system evolution. Moreover, multifractal analysis is effective for understanding the underlying dynamics of any complex system such as pionisation in high energy nucleus-nucleus interactions. To get both qualitative and quantitative idea concerning the multiparticle production mechanism [7] multifractal analysis is expected to be very fruitful. Such a behavior has been observed for vast majority of high energy multiparticle production experiments [8-10].

The investigation of fractal dimension in hadronic multiparticle production was carried out probably first time by Carruthers and Minh [11]. But there was no formalism developed for a systematic fractal study. A systematic approach for the fractal study was suggested by Hwa ($G_q$ moment) [12]. But that $G_q$ moments are found to be influenced by statistical fluctuations especially for the low multiplicity events. Later in order to avoid large statistical fluctuations and exclude low multiplicity events Hwa and Pan [13] proposed a modified $G_q$ moment method introducing a step function, which acts as a filter to the low multiplicity events [14].Also this modified $G_q$ moments suffer from the demerit that they are defined only for positive orders ($q$) and hence it is unable to explore the whole multifractal spectrum. Afterwards a number of techniques [15-18] were developed for the fractal study of multiparticle data. The techniques developed by Hwa [12] and Takagi [15] are the most popular and have been used in many cases [2, 19-21] to analyze multi-pion production process. However, none of these methods can disentangle the dynamical "signal" from the "background". Trending behaviors usually give rise to spurious multifractal effects for the analyzed series. Therefore it is essential to study the intrinsic fluctuations characterizing the dynamical process after filtering out the average trending behavior.

Sophisticated methods have been invented to characterize the actual fluctuations extracted from the average behavior, and the fractal nature of non-stationary time series. These include- detrended fluctuation analysis (DFA) and its variance [22, 23], the wavelet transform [24, 25] based multi-resolution analysis [26, 27], multifractal detrended fluctuation analysis (MFDFA) [28] etc.. DFA technique [22] was developed in order to determine minutely the presence of any long range correlation [22, 29] in a non-stationary series. However, despite a multitude of real-data analyses, a proper detection of the multifractality in the experimental data still presents much difficulty and is not always reliable [30].MFDFA technique [28] is actually a generalization of standard DFA technique for the characterization of multifractal nature of a series. One main reason to employ MFDFA method is to avoid fallacious detection of correlations leading to multi-fractality which are artifacts evolving due to the non-stationarity of the signal. Thus MFDFA is a powerful technique which has been applied successfully to characterize fluctuation in a variety of fields like finance[31-34],medicine [35, 36], natural science[37, 38],solid state physics[39, 40] etc..

The spectrum of references of application of MFDFA technique is not a complete one. Recently, MFDFA method has been applied to analyze the pseudo rapidity and azimuthal angle distribution of the pions produced in Au + Au interactions at 200 GeV/nucleon by Zhang et al. [41]. They studied a sample of only 10 events. Another group, Wang et al.[42], studied the same interactions for UrQMD generated data using the same method. The DFA and MF-DFA methods are also used by Mali et al. [43] to characterize the particle density fluctuation for $^{28}$Si + Ag/Br interactions at 14.5 GeV/nucleon and $^{32}$S + Ag/Br interactions at 200GeV/nucleon. These analyses [41-43] suggest that the MF-DFA approach is a reasonably good technique for the multifractal analysis of multiparticle production process in high-energy nucleus–nucleus (A-A) interactions and hence should be applied for understanding the dynamics of the process.

In this paper we have studied the pseudo rapidity distribution of the pions produced in $^{16}$O-AgBr interactions at 2.1 AGeV and 60 AGeV, $^{12}$C-AgBr & $^{24}$Mg-AgBr interactions at 4.5 AGeV, and $^{32}$S-AgBr interactions at 200 AGeV in the framework of MFDFA. The analysis is expected to reveal a comparative view of the genuine multifractal parameters for interactions initiated by projectiles of various energies and masses.

## Experimental section

The present analysis is performed on the interactions of $^{16}$O beam at 2.1 AGeV & 60 AGeV, $^{12}$C beam & $^{24}$Mg both at 4.5 AGeV and $^{32}$S beam at 200 AGeV with AgBr being the target present in nuclear emulsion.

ILFORD G5 nuclear photographic emulsion plates were irradiated horizontally with a beam of $^{16}$O nuclei at energy of 2.1 AGeV obtained from BEVALAC Berkley [44]. The data for $^{12}$C–AgBr [45] and $^{24}$Mg–AgBr [46] interactions were obtained by exposing NIKFI BR2 emulsion plates to the beams of $^{12}$C and $^{24}$Mg nucleus, of energy 4.5 AGeV at JINR, Dubna, Russia. The data of $^{16}$O–AgBr interactions at 60 AGeV and $^{32}$S–AgBr interactions at 200 AGeV were obtained by exposing the stacks of ILFORD G5 emulsion plates to a beam of $^{16}$O nucleus of energy 60 AGeV and $^{32}$S nucleus of energy 200 AGeV, respectively, using the super proton synchrotron (SPS) at CERN [47]. A Leitz Ortholux microscope with a 10× objective and 25× ocular lens provided with a Brower travelling stage was used to scan the plates of $^{16}$O–AgBr interactions at 2.1 AGeV. A Leitz Metalloplan microscope with a 10× objective and 10× ocular lens provided with a semi-automatic scanning stage has been used to scan the other four interaction plates. Two observers scanned each plate independently so that the biases in detection, counting and measurement could be minimized and consequently the scanning efficiency could be increased. After finding a primary interaction induced by the incoming projectile, the number of secondary tracks in an event belonging to each category was counted using oil immersion objectives. Measurements were carried out with the help of an oil-immersion objective of 100× magnification. The measuring system fitted with both the microscopes has 1μm resolution along the *X*- and *Y*-axes and 0.5μm resolution along the *Z*-axis.

Events were chosen according to the criteria given below.
(a) The incident beam track would have to lie within 3°from main beam direction. (b) Events that occurred within 20 μm from the top and bottom surfaces of the pellicle were rejected. (c) All the primary beam tracks were followed in the backward direction to ensure that the events chosen did not include interactions from the secondary tracks of the other interactions.

According to nuclear emulsion terminology [48], particles emitted after interactions can be classified as the shower, gray and black particles. Shower particles are mostly (about more than 90%) due to pions with a small admixture of K-mesons and hyperons having ionization $I \leq 1.4I_0$ and velocity greater than $0.7c$ where $c$.$I_0$is the minimum ionization of a singly charged particle produced in the emulsion medium. Grey particles are mainly fast target recoil protons with energies up to 400 MeV. They have ionization $1.4I_0 \leq I < 10I_0$. They have velocities lying between $0.3c$ and 0.7$c$. Black particles consist of both singly and multiply charged fragments of the target nucleus with ionization $I \geq 10 I_0$ and velocity less than 0.3$c$.

Along with the above stated three kinds of particle, there could also be a few projectile fragments. These projectile fragments are the spectator parts of the incident projectile nuclei that do not directly participate in an interaction. In the experiments with the nuclear emulsion track detectors, interactions may be with three different type of targets e.g., hydrogen (H), light nuclei (CNO) and heavy nuclei (AgBr) present in the emulsion medium. Events, with $N_h \leq 1$, occur because of the collision between hydrogen and the projectile beam. Events with $2 \leq N_h \leq 8$ are due to collisions of projectile with light nuclei and events with $N_h > 8$ are due to collisions with heavy nuclei. Here $N_h$, the number of heavy tracks, is the total number of black and gray tracks. In our study, only events having a number of heavy tracks greater than 8 ($N_h > 8$) have been selected to exclude the H and CNO events.

According to the selection procedure mentioned above, we have chosen 730 events of $^{16}$O–AgBr interactions at 2.1AGeV [44], 800 events of $^{12}$C–AgBr [49] and $^{24}$Mg–AgBr [50] interactions at 4.5AGeV, 250 events of $^{16}$O–AgBr interactions at 60AGeV [51] and 140 events of $^{32}$S–AgBr interactions at 200AGeV [52]. The present analysis has been performed on the pion tracks only. The emission angle ($\theta$) was measured for each pion track with respect to the beam direction by taking readings of the coordinates of the interaction point ($X_0$, $Y_0$, $Z_0$), coordinates ($X_1$, $Y_1$, $Z_1$) of a point at some distance away from the interaction point on each secondary track and coordinates ($X_i$, $Y_i$, $Z_i$) of a point on the incident beam. From '$\theta$' the pseudo-rapidity variable ($\eta = -\ln \tan \frac{\theta}{2}$), which may be treated as a convenient substitute of the rapidity variable of a particle when the rest mass of the particle can be neglected in comparison to its energy or momentum, was calculated for each pion track.

## MFDFA Method

The MFDFA technique was developed by Kantelhardt et al. [28] as a generalization of standard DFA method to analyze non-stationary time series. If {$x_k$, k=1, 2, 3…, N} be the signal of length N, MFDFA consists of the following steps [28] among which the first three steps includes the ordinary DFA technique.

Step 1.  Calculation of the signal profile, which is the cumulative sum of the signal to be analyzed, according to the equation (1)

$$y(j) = \sum_{k=1}^{j} [x_k - \langle x \rangle]$$, where j runs from 1 to N ............(1)

Step 2.  Division of the profile $y(j)$ into $N_s = \text{int}(N/s)$ numbers of non-overlapping segments of length $s$. Since N is not an integer multiple of $s$, a small part of the signal will be left at the end. In order to include that part the same process is repeated starting from the other end. Thus one obtains altogether $2N_s$ number of segments.

Step 3. Determination of the local trend associated with each of such $2N_s$ segments by a least square polynomial fit of the series in a particular segment. And calculation of the variance of the series. The variance of the series relative to the local trend in a given segment v of length s can be calculated as

$$F^2(v,s) = \frac{1}{s}\sum_{j=1}^{s}\{y[(v-1)s+j] - y_{fit}^{lv}(j)\}^2$$

..............(2)

where $y_{fit}^{lv}(j)$ is the fitting polynomial of order *l* for the $v^{th}$ segment.

Step 4. The $q^{th}$ order fluctuation function $F_q(s)$ is then given by [30] the equation (3)

$$F_q(s) = \left\{\frac{1}{2N_s}\sum_{v=1}^{2N_s}[F^2(v,s)]^{q/2}\right\}^{1/q} \quad \text{for } q \neq 0$$

$$= exp\left\{\frac{1}{4N_s}\sum_{v=1}^{2N_s}\ln F(v,s)\right\} \quad \text{for } q = 0$$

..........(3)

Step 5. To investigate the scaling behavior one has to calculate $F_q(s)$ for various *s* values. It should be remembered that the sample size of the smallest segment (or scale) should be much larger than the polynomial order *l* in order to prevent an over fitted trend.

**Scaling behavior and multifractality:**

Presence of any long range correlation in the system will manifest in the multifractal behavior of the system and then $F_q(s)$ will show a power law dependence on scale *s* like

$$F_q(s) \propto s^{h(q)}$$

..............(4)

The scaling exponent $h(q)$ is known as generalized Hurst exponent. For q=2 it reduces to the ordinary Hurst exponent H [53]. $H \neq 0.5$ indicates the presence of long range correlation in the system. An exponent $H > 0.5$ corresponds to positive correlation in the system and H<0.5 corresponds to the presence of anti-correlation in the system. For a mono-fractal series $h(q)$ is independent of q and the fluctuation function $F_q(s)$ is similar for all sub-signals. On the other hand if large and small fluctuations scale differently then there will be a notable dependence of $h(q)$ on q. Particularly for multifractal series $h(q)$ decreases with order q. for +ve *q* values h(q) describes the scaling behavior of the segments with large fluctuations and for –ve *q*'s it describes the scaling behavior of the segments with small fluctuations.

The multifractal exponent $\tau(q)$ can be derived from $h(q)$ [28, 53] using equation (6)

$$\tau(q) = qh(q) - 1 \quad \ldots(5)$$

The singularity spectrum $f(\alpha)$ can be obtained from $\tau(q)$ through a Legendre transform [18, 53, 54]:

$$f(\alpha) = q\alpha - \tau(q)$$

$$\text{where } \alpha = \partial\tau(q)/\partial q \quad \ldots(6)$$

The width of the multifractal spectrum is the difference between the maximum and minimum value of $\alpha$ i.e.

$$\text{width of spectrum} = \alpha_{max} - \alpha_{min} \quad \ldots(7)$$

It has been proposed by Y. Ashkenazy et al. [55] and Shimizu *et al.*[56] that the width of a multifractal spectrum is a measure of the degree of multifractality. Broader the spectrum richer the multifractality [57].

Though originally MFDFA technique was developed for non-stationary time series analysis but it can also be used equally for non-uniform distribution like rapidity distribution of high energy multiparticle production process [41-43]. In principle multifractality of a natural system may originate either (a) due to broad probability density function (non-Gaussian distribution) of the concerned parameter, or (b) due to the simultaneous presence of dissimilar characteristics of long range correlations for large as well as small fluctuations, or sometimes (c) due to both of reason (a) and (b) [58]. MFDFA technique allows one to identify the source of multifractality when it is applied to the randomly shuffled distribution of the same. For shuffled distribution though the probability distribution function remains unchanged but all possible origin of correlation is wiped out from the distribution. Thus if multifractality is originated from type b) then randomly generated distribution will show an arbitrary behavior with $h(q)_{random}=0.5$. On the other hand if the source of multifractality is type a) then for the generated distribution generalized Hurst exponent will be identical to that of the original one ($h(q)_{random}= h(q)_{experimental}$). Furthermore for the reason c) the generated series will show a weaker multifractality than that of the experimental one.

## Results and Discussions

In the current analysis we have focused on $^{16}$O-AgBr interactions at 2.1AGeV & 60AGeV, $^{12}$C-AgBr & $^{24}$Mg-AgBr interactions at 4.5AGeV, and $^{32}$S-AgBr interactions at 200AGeV. For each interaction we have considered only those events which have multiplicities greater than average multiplicity of the corresponding interaction. Pseudo rapidity distribution of the pions produced in each such event was obtained for 0.1 rapidity interval. MFDFA technique as discussed has been utilized to analyze the pseudo rapidity distribution of the shower particles corresponding to each selected events. Here the term scale(*s*) corresponds to the width of each segment in the pseudo rapidity space. The smallest scale is so chosen that at least 10 data points are within it i.e. it covers one pseudo rapidity unit in the pseudo rapidity space. The fluctuation function $F_q(s)$ has been calculated for different

's' values with 'q' varying from -5 to +5 in steps of 1 including 0 using equation (3) for each selected events. Event average of fluctuation function ($<F_q(s)>$) is obtained by averaging over all selected events. The plot of $\log <F_q(s)>$ with log(s) for q=+5, 0 and -5 is shown in figure 1 for all interactions. For any particular order $q$ the event-average fluctuation function $<F_q(s)>$ manifests power law dependence on $s$ consequently all the plots are of straight line nature with different slope. The point symbols in figure 1 represent the values of $\log <F_q(s)>$ for different $\log(s)$ whereas the solid lines represent the regression fit of the same. $<F_q(s)>$ for different 'q' values shows a convergent trend as the scale increases. The slope for linear fit of $\log <F_q(s)>$ vs. $\log s$ graph provides the generalized Hurst exponent $h(q)$. $h(q)$ values vary for different $q$'s.

The variation of $h(q)$ with q has been depicted in figure 2. For all the interactions $h(q)$ is $q$ dependent and it decreases with $q$, signifying multifractal nature of the rapidity distribution as manifested by the different scaling behaviors for large and small fluctuations. $h(q)$ values for q<0 seem to be higher than those for q>0. This typical feature of $h(q)$, is consistent with the observation by Kantelhardt et al. [27] for multifractal time series. The ordinary Hurst exponent H i.e., $h(q)$ for q=2 are given in table 1 for all the interactions. H, as mentioned earlier, gives an idea about the correlation present in the system. For all the considered interactions H>0.5 indicates that long range correlation exist in our data for all the interactions.

We have also calculated $\tau(q)$ for different $q$ values according to equation (5). The plot of $\tau(q)$ against $q$ is shown in figure 3. The non-linear variation of $\tau(q)$ with $q$ also confirms the fact that the pion density fluctuation in the rapidity space is multifractal in nature i.e. pion production mechanism has an inbuilt multi-scale self-similarity property.

In order to shed light on the possible origin of the demonstrated multifractal behavior we have randomly shuffled the rapidity distribution of each selected event for all interactions and have employed MFDFA method to the shuffled data using the same approach as before. Results found to be identical for all the interactions. Here we present for $^{32}$S-AgBr interactions only.

Figure 4 shows the plot of $h_{q, shuffle}$ with $q$ along with $h_{q, expt.}$ vs. $q$. $h_{q, shuffle}$ values remain around 0.5 (average 0.56) with a minor dependence on $q$ with $h_q$ (q=2)=0.51. H≈0.5 suggests that multifractality in our data is mainly sourced from the presence of long range correlation in the data. $\tau(q)$ vs. $q$ plots (figure 5)for the shuffled and experimental data further strengthens this fact . $\tau(q)_{,shuffle}$ shows an almost linear behavior instead of non-linear one as for the experimental data. Similar behaviors were observed for the other interactions also.

To have a quantitative idea about the multifractality we have computed the multifractal spectrum corresponding to each interactions. The multifractal spectrums corresponding to different interactions have been displayed in figure 6. Figure 7shows the spectrum for $^{32}$S-AgBr interactions both for expt. and shuffled distributions. For shuffled data spectrum has negligible width ($\alpha$=0.38) w.r.t. experimental data ($\alpha$=2.39) and also the spectrum shifted to the left of the experimental spectrum as expected when multifractality is mainly sourced [58] from type (b) cause as discussed earlier.

To compare the strength of multifractality we have calculated the width of the multifractal spectrum for each of the interactions. These values have been presented in Table 2. The spectrum width is found to vary for different interactions suggesting that the strength of multifractality depends upon the mass and energy of the projectile beam. The spectrum width values reveal the following notable features of the heavy ion interaction process

- Comparison of $^{16}$O-AgBr interactions at 2.1 AGeV & at 60 AGeV indicates that for a particular projectile, strength of multifractality increases with the energy of the projectile.
- From $^{24}$Mg-AgBr interactions & $^{12}$C-AgBr interactions at same energy 4.5 AGeV, it is evident that the strength of multifractality decreases with projectile mass.
- For projectiles with different mass and energies multifractality decreases if there is a simultaneous decrease in energy and increase in mass of the projectile beam (e.g. the pair of interactions i) $^{16}$O-AgBr at 60 AGeV & $^{24}$Mg-AgBr at 4.5 AGeV, ii) $^{12}$C-AgBr at 4.5 AGeV & $^{16}$O-AgBr at 2.1 AGeV). This observation agrees very well with that of the previous two observations.
- If projectile mass and energy simultaneously changes in a same manner, i.e. either both increase or both decreases, the strength of multifractality is predominantly affected by the mass of the projectile if energy is of the same order. This fact is revealed if we consider the pair of interactions i) $^{16}$O-AgBr at 2.1 AGeV & $^{24}$Mg-AgBr at 4.5 AGeV, ii) $^{16}$O-AgBr at 60 AGeV & $^{32}$S-AgBr at 200 AGeV. On the other hand, if mass difference is little, strength of multifractality is mainly determined by the energy difference of the projectile. This is demonstrated by the interactions $^{12}$C-AgBr at 4.5 AGeV & $^{16}$O-AgBr at 60 AGeV pair.
- Finally if we go from relativistic energy regime to the ultra relativistic one, influence of energy is much more prominent than mass. This is noticed comparing $^{16}$O-AgBr at 2.1 AGeV, $^{24}$Mg-AgBr at 4.5 AGeV, $^{12}$C-AgBr at 4.5 AGeV interactions with the $^{32}$S-AgBr interactions at 200 AGeV.

## Conclusions

We have presented a systematic study on pion density fluctuation in pseudo rapidity spectra in the framework of sophisticated MFDFA technique using various heavy ion projectiles covering a wide range of energy starting from a few GeV to a few hundred GeV. The main observations of the analysis may be summarized as follows:

- Both the h(q) and τ(q) spectra speaks in favor of multifractal pion density fluctuation in pseudo rapidity space for all the considered A-A interactions.
- Multifractal parameter $h_q$ and $τ_q$ and the multifractal spectrum for the shuffled data clearly demonstrate that the observed multifractality of the experimental data does not originate from the trivial broad probability density distribution but occurs due to dynamics.
- The study suggests multifractal pion production dynamics in heavy ion interactions.
- The strength of multifractality is influenced both by mass and energy of the projectile beam.
- For the same projectile beam multifractality increases with projectile energy. On the other hand it decreases with mass of the projectile beams having same energy.

- When mass and energy both changes, effect of mass is dominant if their energy does not differ much. But multifractality is much more influenced by energy of the projectile beam if mass difference is small.
- From the value of ordinary Hurst exponent we can conclude that long range correlation is present among the pions for all of the considered heavy ion interactions irrespective of projectile mass and energy.

## Acknowledgement

The authors are grateful to Prof. P.L Jain, State University of Buffalo, Buffalo, U.S.A, Prof. P.S. Young of Mississippi State University, Mississippi and Prof. K.D. Tolstov, JINR, Dubna, U.S.S.R. for providing them exposed and developed emulsion plates used for this analysis. One of Us Miss Gopa Bhoumik acknowledges CSIR for a Junior Research Fellowship (Sanction No: 09/096(0802)/2013-EMR-I) during this work.

## References:

[1] Mandelbrot B.B., 1999 The Fractal Geometry of Nature, Freeman, San Francisco.

[2] Shafiq Ahmad et al 2006 *J. Phys. G: Nucl. Part. Phys.* **32** 1279

[3] Mohsin Khan M et al. 2011 *Indian J. Phys.* **85(1)** 189-193

[4] Singh G et al 1994 *Physical Review C* **50.5**2508

[5] Bershadskii A 2000 *Journal of Physics G: Nucler and Particle Physics* **26.7** 1011

[6] Chen Z et al. *Phys. Rev.* E **65**(2002)041107-041122

[7] Chekanov S V et al 1996 *Journal of Physics G :Nucl. Part. Phys.***22** 601-610

[8] Bialas A 1991 *Nucl. Phys. A* **525** 345

[9] Peschanski R 1991 *Journal of Modern Phys. A***21**3681

[10] E. A. De Wolf, I. M. Dremin, W. Kittel, *Scaling Laws for Density Correlations and Fluctuations in Multiparticle Dynamics* preprint HEN-362(93) IIHE-93.01 FIAN/TD-09/93 (update 1995), hep-ph/9508325, Phys. Rep. (in print)

[11] P. Carruthers and Minh Duong-Van, Los Alamos Report No. LA-UR-83-2419 (unpublished).

[12] Hwa R C 1990 *Phys. Rev.* D **41** 1456–1462

[13] Hwa R C et al 1992 *Phys. Rev.* D **45** 1476

[14] Dermin I M 1988 *Mod. Phys. Lett.* A **3** 1333

Carruthers P 1989 *Int. Journal of Mod. Phys.* A **4** 5587

BraxPh and Peschanski R 1990 *Nucl. Phys.* B**346**650

[15] Takagi F 1994 *Phys. Rev. Lett.***72**32–35

[16] Paladin G et al 1987 *Phys. Rep.***156**147–225

[17] Grassberger P et al 1984 *Physica* D **13**34–54

[18] Halsey T C et al 1986 *Phys. Rev.* A **33** 1141–1151

[19] Haque M I et al 2014 *Journal of Modern Physics* **5**1 889

[20] Ahmad N et al 2014 *Journal of Modern Physics* **5** 1288

[21] Ghosh D et al 1998 *Phys. Rev.* C **58** 3553

[22] Peng C K et al 1994*Phys. Rev.* E **49** 1685

[23] Kantelhardt J W *et al* 2003 *Physica* A **330** 240

[24] I. Daubechies, Ten Lectures on Wavelets, SIAM, Philadelphia, 1992

[25] S. Mallat, A Wavelet Tour of Signal Processing, *Academic Press*, 1999

[26] Manimaran P *et al* 2005 *Phys. Rev.* E **72** 046120

[27] Manimaran P *et al* 2006 *J. Phys.* A **39** L599

[28] Kantelhardt J W *et al* 2002 *Physica* A **316** 87

[29] Taqqu M S *et al* 1995 *Fractals* **3** 785

[30] Bouchaud J P *et al* 2000 *Eur. Phys. J.* B **13** 595

[31] Mali *P et al* 2014 *Physica* A **413** 361 and the references therein

[32] Yuan Y *et al* 2009 *Physica* A **388** 2189

[33] Wang Y *et al* 2009 *Int. Rev. Financ. Anal.* **18(5)** 271

[34] Norouzzadeh P 2006 *Physica* A **367** 328

[35] Zhou P *et al* 2007 Woeld Congress on Medical Physics and Biomedical Engineering, Springer, Berlin, Heidelberg, New York

[36] Dutta S *et al* 2013 *Front. Physiol.* **4** 274.

[37] Mali P, Multifractal characterization of global temperature anomalies, Theor. Appl. Climatol., http://dx.doi.org/10.1007/s00704-014-1268-y.

[38] de Benicio R B *et al* 2013 *Physica* A **392** 6367

[39] Kantelhardt J W *et al* 1999 *Physica* A **266** 461

[40] Vandewalle N *et al* 1999 *Appl. Phys. Lett.* **74** 1579

[41] Zhang Y X *et al* 2007 *Internat. J. Modern Phys.* A **23(18)** 2809

[42] Wang X *et al* 2013 *Internat. J. Modern Phys.* E **22(4)** 1350021

[43] Mali P *et al Physica* A **424** 25

[44] Ghosh D *et al* 1994 *Phys. Rev.* C **49** R1747

[45] Ghosh D *et al* 1989 *Mod. Phys. Lett.* A **4** 1197
[46] Ghosh D *et al* 1987 *Nucl. Phys.* A **468** 719
[47] Sengupta K *et al* 1990 *Phys. Lett.* B **236** 219

[48] Powell C F *et al* 1959 *Oxford: Pergamon* pp 450–64 and references there in

[49] Ghosh D *et al* 1994 *J. Phys. G: Nucl. Part. Phys.* **20** 1077
[50] Ghosh D *et al* 1997 *Phys. Rev.* C **56** 2879
[51] Ghosh D *et al* 1999 *Phys. Rev.* C **59** 2286
[52] Ghosh D *et al* 1995 *Phys. Rev.* C **52** 2092

[53] J. Feder 1988 Fractals, Plenum Press, New York

[54] Peitgen H.-O. *et al* 1992 Chaos and Fractals, Springer, New York (Appendix B)

[55] Ashkenazy Y *et al* 2003 *Geophysical Research Letters* **30** 2146

[56] Shimizu Y *et al Fractals* **10** (2002) 103-116

[57] Telesca L *et al Phys Chem Earth* **29** (2004) 295-303

[58] Barman C *et al Journal of Earthquake Science: Nat Hazards* DOI 10.1007/s11069-015-1747-1

## Figure captions:

**Figure 1.** Plot of logarithm of fluctuation function with logarithm of scale for (a) $^{12}$C-AgBr interactions at 4.5AGeV, (b)$^{24}$Mg-AgBr interactions at 4.5AGeV(c)$^{16}$O-AgBr interactions at 2.1AGeV, (d)$^{16}$O-AgBr interactions at 60AGeV, (e)$^{32}$S-AgBr interactions at 200AGeV.

**Figure 2.** Plot of variation of generalized Hurst exponent ($h_q$) with order (q) for all the interactions

**Figure 3.** Plot of $\tau(q)$ vs. *q* for different interactions

**Figure 4.** Comparison of ordinary Hurst exponent for experimental and shuffled data for $^{32}$S-AgBr interactions at 200AGeV

**Figure 5.** Comparison of $\tau_q$ spectra for experimental and shuffled data for $^{32}$S-AgBr interactions at 200AGeV

**Figure 6.** Multifractal Spectrum for the considered interactions

**Figure 7.** Comparison of multifractal spectrum for experimental and shuffled data for $^{32}$S-AgBr interactions at 200AGeV

## Table captions:

**Table 1.** Hurst exponent for the different interactions

**Table 2.** Width of multifractal spectrum for different interactions:

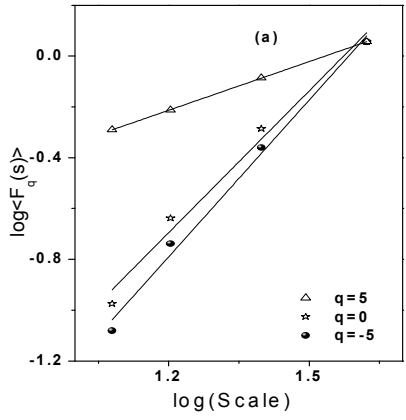
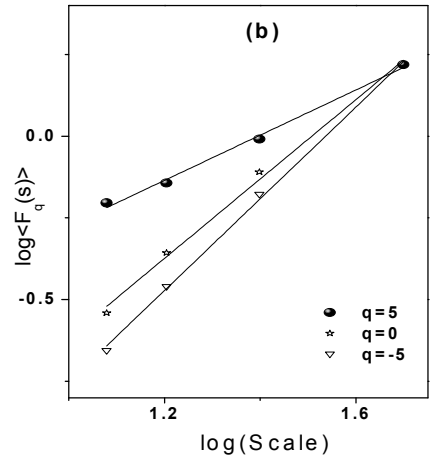
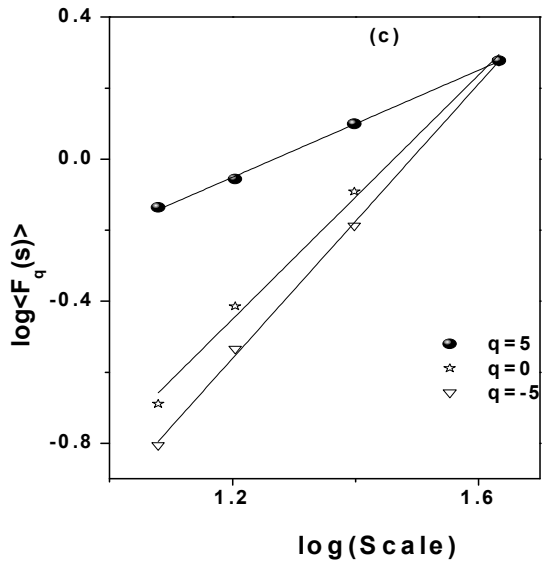
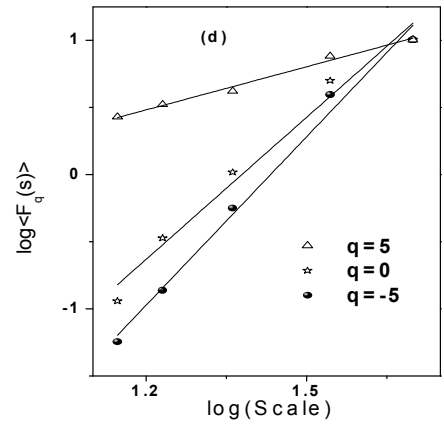
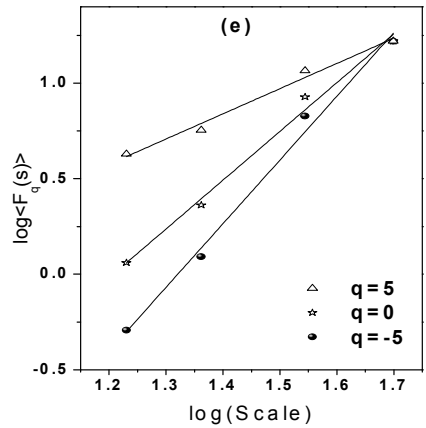

Figure 1.

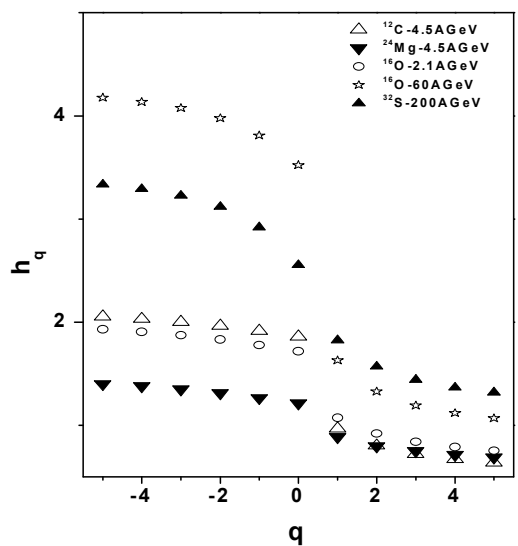

Figure 2.

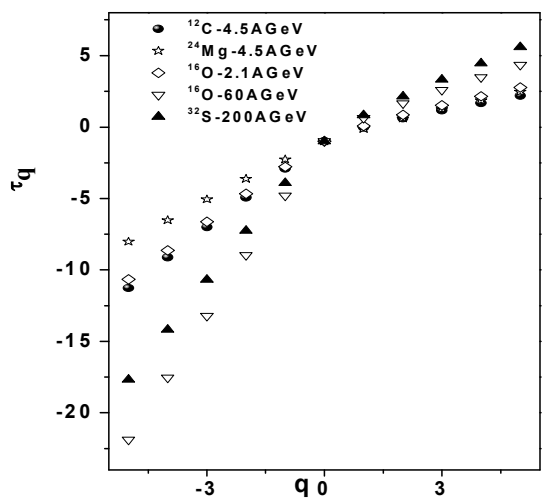

Figure3.

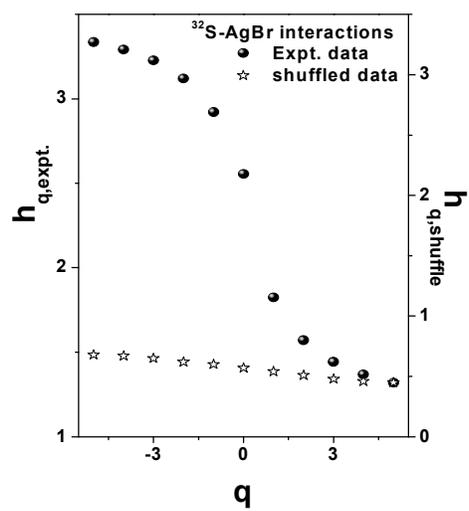

Figure 4

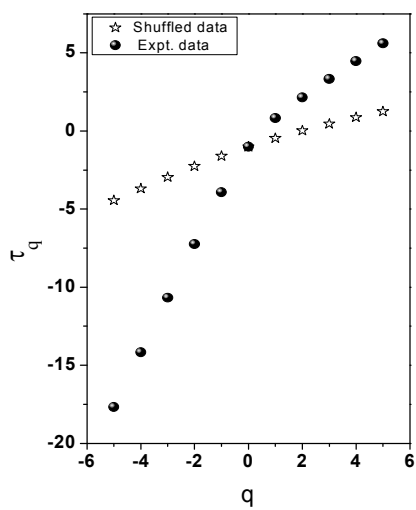

Figure 5

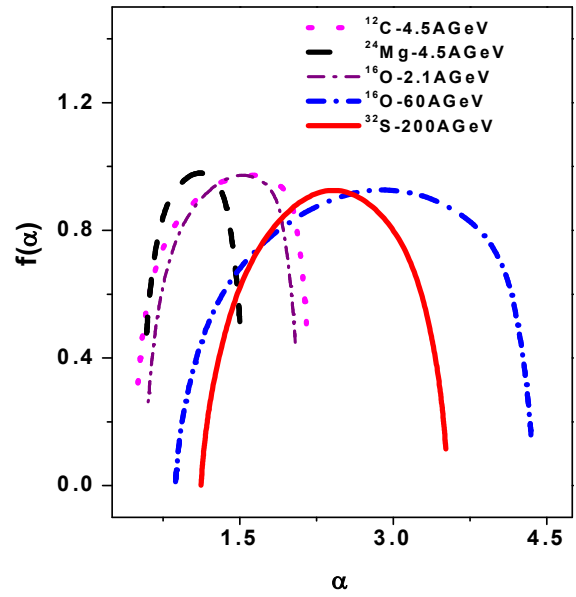

Figure 6.

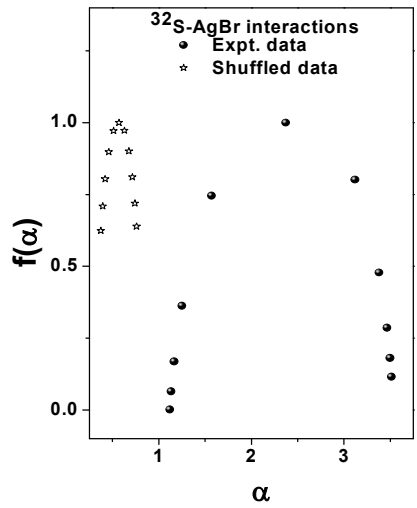

Figure 7.

Table 1.

| Interactions | Value $h_q$ at q=2 |
|---|---|
| $^{12}$C-AgBr interactions at 4.5AGeV | 0.803 |
| $^{16}$O-AgBr interactions at 2.1AGeV | 0.922 |
| $^{16}$O-AgBr interactions at 60AGeV | 1.352 |
| $^{24}$Mg-AgBr interactions at 4.5AGeV | 0.801 |
| $^{32}$S-AgBr interactions at 200AGeV | 1.571 |
| Shuffled $^{32}$S-AgBr interactions at 200AGeV | 0.511 |

Table 2.

| Interactions | $\alpha_{max}$ | $\alpha_{min}$ | Width of spectrum |
|---|---|---|---|
| $^{12}$C-AgBr interactions at 4.5AGeV | 2.15 | 0.49 | 1.66 |
| $^{16}$O-AgBr interactions at 2.1AGeV | 2.04 | 0.60 | 1.44 |
| $^{16}$O-AgBr interactions at 60AGeV | 4.34 | 0.87 | 3.47 |
| $^{24}$Mg-AgBr interactions at 4.5AGeV | 1.50 | 0.58 | 0.92 |
| $^{32}$S-AgBr interactions at 200AGeV | 3.51 | 1.12 | 2.39 |
| Shuffled $^{32}$S-AgBr interactions at 200AGeV | 0.76 | 0.38 | 0.38 |